\documentclass[apjl]{emulateapj}

\def\ltsima{$\; \buildrel < \over \sim\;$}
\def\ltsim{\lower.5ex\hbox{\ltsima}}
\def\gtsima{$\; \buildrel > \over\sim \;$}
\def\gtsim{\lower.5ex\hbox{\gtsima}}
\def\ms{$M_{\odot}$ }
\def\msp{$M_{\odot}$}

\slugcomment{Accepted for publication in ApJ Letters}

\begin{document}
\title{A new contributor to chemical evolution in high-redshift galaxies}

\author{ Takuji Tsujimoto$^1$}

\altaffiltext{1}{National Astronomical Observatory, Mitaka-shi,
Tokyo 181-8588, Japan; taku.tsujimoto@nao.ac.jp}

\begin{abstract}
The recent discovery of a new population of stars exhibiting unusual elemental abundance patterns characterized by enhanced Ti to Ga elements and low $\alpha$ and $n$-capture elements suggests the contribution of a new class of supernovae, probably a kind of Type Ia supernovae associated with close binary evolution. The role of these supernovae in chemical evolution is negligible in normal galaxies that undergo moderate star formation such as our own. Thus, while the frequency of occurrence would be too low to detect in low-redshift galaxies, it may represent a prominent population in high-redshift objects such as early epoch massive elliptical galaxies and QSOs. The chemical contributor of this proposed type of supernovae in combination with recognized supernovae is shown to be compatible with the recent observational features in the distant universe, successfully reproducing the Type II supernovae-like abundance pattern with enhancement of Ga and Ge in the gas of  newborn massive galaxies and high iron abundances in QSOs even at redshifts of around 6. 
\end{abstract}


\section{Introduction}

Accumulated observational results of the distant universe have revealed very early chemical enrichment in high-redshift quasi-stellar objects (QSOs). The FeII/MgII emission line ratio of QSOs represents a good tracer for this study, and \citet{Iwamuro_02} and \citet{Dietrich_03} found that the FeII/MgII ratios at redshifts $z$ of $\sim 5$ are comparable to those of low-$z$ QSOs ($\sim 4-5$), in some cases being even higher~\citep{Iwamuro_02}. Surprisingly, similar ratios have also been found for QSOs at $z\sim 6$~\citep{Freudling_03, Maiolino_03}. \citet{Maiolino_03} have shown that QSOs at $z\sim 6.3$ exhibit FeII/MgII ratios about twice as high as those of low-$z$ QSOs. Although the correlation between the FeII/MgII line ratio and the actual Fe/Mg abundance ratio still remains uncertain \citep[see][]{Verner_99}, these high-redshift QSOs are likely to be enriched in Fe from Type Ia supernovae (SNe Ia) as in low-$z$ QSOs \citep{Matteucci_01, Dietrich_03, Freudling2_03, Maiolino_03}. This follows from the firm bounding of the Fe II/Mg II line ratio in theoretical solar abundance models between 1.5 and 4~\citep{Wills_85}. However, this implication is at odds with the fact that a cosmology of $H_0=71$~km~s$^{-1}$~Mpc$^{-1}$, $\Omega_\Lambda=0.73$, and $\Omega_m=0.27$ \citep{Bennett_03} results in an age of less than 1~Gyr for QSOs with $z\sim 6.3$.  At such an early stage, it appears to be impossible for SNe Ia to enrich the interstellar matter (ISM) at a high Fe abundance.

Such early enrichment in QSOs may reflect the earliest SNe Ia, which may have had different characteristics from those of the solar neighborhood of SNe Ia. In addition, even in the solar neighborhood, some relics implying the existence of these SNe Ia might be embedded in the stellar abundances of metal-poor stars due to their short lifetime.  Recently, \citet{Ivans_03} found two stars belonging to a new class of low-$\alpha$ stars that exhibit enhanced abundances of elements from Ti to Ga, features never seen in other low-$\alpha$ stars. The deficiency of $\alpha$-elements is significant in these stars, with characteristics such as [Mg/Fe]~=~$-0.64$ and [Si/Fe]~=~$-0.97$, whereas other low-$\alpha$ stars are only mildly deficient with [$\alpha$/Fe]$\sim 0$ at most. Such elemental abundance features are reminiscent of the nucleosynthesis of SNe Ia. Recent studies on the chemical compositions of metal-poor stars have revealed that these stars might have inherited the abundance pattern of the ejecta of the preceding single SN~\citep[e.g.,][]{Shigeyama_98}. If this is the case, these two stars may have been born from the ejecta of some kind of SN Ia, which would have had an explosion timescale comparable to those of massive stars ending in Type II SNe (SNe II). Otherwise, as in the general cases of normal SNe Ia, the fossil imprint of nucleosynthesis in SNe Ia will not be retained in stellar abundances due to mixing with the products from SNe II within the ejecta.

The merging of double white dwarfs (WDs) as the end result of a close binary consisting initially of two massive stars of $\sim 8$\ms could produce SNe Ia with an explosion timescale of the order of $10^6$--$10^7$~yrs~\citep{Iben_84}.  In this case, the combined mass of merging WDs is likely to be super-Chandrasekhar~\citep{Tutukov_94}. Although the photometric and spectroscopic features of most SNe Ia in the local universe can be successfully reproduced using Chandrasekhar mass models~\citep[e.g.,][]{Hoflich_96}, some implications of a super-Chandrasekhar explosion are present for a peculiar supernova SN1991T~\citep{Fisher_99}. Theoretically the issue of whether a super-Chandrasekhar coalescence can lead to a SN Ia explosion is an unresolved question \citep[see][]{Branch_95}.  In any event, such SNe Ia are, if indeed they exist, expected to be locally rare events. However, the situation could be significantly different in other galaxies because of the potential for rates of close binary system formation in high-density environments, as suggested by the recent globular cluster (GC) study~\citep{Pooley_03}, which revealed with high confidence that the number of close binaries in GCs increases with the stellar encounter rate of the cluster. Therefore, it is likely that the large amount of Fe observed in high-$z$ QSOs might be attributable to numerous SNe Ia originating from such close binaries.

Other signature of these potential SNe Ia can be seen in the high-$z$ absorption system of QSOs. Recently, \citet{Prochaska_03} observed elemental abundances in a damped Lyman alpha (DLA) system at $z=2.626$. They concluded that this DLA galaxy is the progenitor of a massive elliptical galaxy based on the [$\alpha$-elements/Zn] ratios, which are suggestive of an SN II origin \citep[see also,][]{Fenner_04} and imply that the potential age of this galaxy is of the order of several $10^8$~yrs. However, this conclusion presents the puzzling problem of how the Ge abundance could become so enhanced in such a young galaxy \citep{Prochaska2_03}. Incidentally, among the enhanced elements for the two solar neighborhood stars in question, a remarkable enhancement is seen in the Ga abundance. Since Ga and Ge may be produced in a similar fashion, the Ge abundance in this galaxy could be derived from the proposed new class of SNe Ia. This consideration is appropriate from the viewpoint that in the formation of massive elliptical galaxies, the gas density is expected to be high, leading to a large fraction of close binary systems.

This correspondence discusses three mysteries in the elemental abundance features of recently observed solar neighborhood stars, a DLA galaxy at $z=2.626$, and high-$z$ QSOs of around $z=6$ in the framework of a unified scheme that considers a contribution from a new class of SNe Ia. It should be, 
however, stressed that the proposed scenario is based on a number of speculations that involve large 
uncertainties and are going to be investigated more deeply point by point by future works.

\section{Fossil Imprints in Metal-Poor Stars in the Solar Neighborhood}

The two stars in question, G4-36 and CS22966-043, were discovered by \citet{Ivans_03}. Deficiencies of $\alpha$-elements and enhancements of Fe-group elements and Ga with respect to Fe in these stars are conspicuous, particularly in CS22966-043, suggesting that these stars must represent the first few generations of stars, as the high deficiency/enhancement would be reduced in later generation stars due to a gradual increase in the contribution from the ISM with high/low [X/Fe] ratios. 
However, we need a caution against a highly enhanced Ga abundance, as mentioned by \citet{Ivans_03} that there exists the possibility that the detected feature in their spectra is a result of 
some other unidentified element and not due to enhanced Ga.  Contrary to these enhanced elements, 
a severe deficiency in Cu is seen in one of two stars, which seems incompatible with the enhancements of other elements.
These two stars also exhibit low [$n$-capture (Sr, Ba)/Fe] ratios. These [Ba/Fe] ratios are comparable to those for very metal-poor stars in the range of $-4<$[Fe/H]$<-3$, which means that the abundances of $n$-capture elements in these stars originate from the ISM containing the $r$-process products from massive stars during a very early epoch~\citep{Tsujimoto_00}, whereas \citet{Ivans_03} concluded  from a special attention to the Sr abundances that the origin of these $n$-capture elements is not a $r$-process-only site. However, the Sr abundance has an enormous scatter among low-metallicity stars \citep{Travaglio_04}, and is not therefore appropriate to  the $s$-/$r$-process indicator. In any event, the definitive conclusions on the origin of $n$-capture elements will be clarified by the acquaintance of the $s$-/$r$-process indicators such as  the [Ba/Eu] or [La/Eu] ratios. It is also of note that these two stars have similar metallicities [Fe/H]$\sim -2$ as well as similar abundance patterns. Taking the mass of gas swept up by the SN remnant ($\sim 6.5\times10^4$\msp)~\citep{Shigeyama_98}, the metallicity [Fe/H] of $\sim -2$ corresponds to the Fe mass $\sim 0.8$\ms from the preceding SN. This Fe mass is comparable to that synthesized in SNe Ia. Otherwise, a hypernova may be a candidate for the preceding SN, considering the features of the well-known hypernova SN1998bw, which ejected $\sim 0.4-0.7$\ms of Fe~\citep{Nakamura_01}. However, theoretical calculations on hypernovae~\citep{Umeda_02,Maeda_03} predict totally different abundance patterns from those observed for the present two stars. In addition, the estimated mass $\sim$0.8\ms of Fe is compatible 
with the Fe mass inferred from the luminosity of SN1991T which is brighter by $\sim0.3$ mag than the 
mean magnitude of other observed  SNe Ia \citep{Ritcher_01}. Therefore, from the elemental abundance features and estimated mass of Fe, these stars are likely to have formed from the ejecta of a kind of SNe Ia.

The proposed SNe Ia are considered to differ from normal SNe Ia in that (i) the timescale for the explosion is comparable to that for massive stars, and (ii) [Ti$-$Ni/Fe] ($\sim 0.3-0.5$) and [Zn$-$Ge/Fe] are much higher. The first property might be possible as discussed in the previous section, though whether such a short timescale for the explosion through an approach of double WDs by the gravitational energy loss and subsequently the formation of an accretion disk would be realized is still an open question. Further, the second property is more debatable. According to detailed calculations of nucleosynthesis in Chandrasekhar mass models for SNe Ia~\citep{Iwamoto_99}, these enhancements are incompatible with  the present SNe Ia models. For example, the predicted [Mn/Fe] is nearly solar, whereas these two stars exhibit [Mn/Fe] ratios of $\sim 0.4$.  More seriously, these models produce little Zn, Ga and Ge. The predicted [Zn/Fe] ratios in the various models are around $-1$ to $-2$, whereas the present two stars exhibit [Zn/Fe] ratios of $\sim +1$. This gives rise to the question if this large difference can be diminished by other SN Ia models such as super-Chandrasekhar mass models. Although  the answer remains unknown at present, it should be noted that the current Chandrasekhar mass models for SNe Ia have already encountered some problems. Recent observations of [Zn/Fe] for solar neighborhood stars~\citep[e.g.,][]{Nissen_04, Cayrel_04} have revealed a clear trend of [Zn/Fe] against [Fe/H], where [Zn/Fe] exhibits plateau-like behavior ($\sim$ 0.1) between [Fe/H]~$=-2.5$ and 0, and gradually increases with decreasing [Fe/H] at [Fe/H]$< -2.5$. There is essentially no change in [Zn/Fe] for [Fe/H]$> -2.5$, implying that nucleosynthesis in normal SNe Ia should give [Zn/Fe]$_{\rm Ia} \sim 0.1$ as well as the case of SNe II as an average, which is higher by one or two orders of magnitude than those predicted by the current SN Ia models. Furthermore, abundance studies for the solar neighborhood claimed that a substantial fraction of Zn is produced in SNe Ia \citep{Raiteri_92, Matteucci_93}, though it has not ever been confirmed by nucleosynthesis calculations. 
Expecting some progress on this topic, the existence of a new class of SN characterized by short lifetime and enhanced [Ti$-$Ge/Fe] ratios is assumed here as a potentially extreme minority population of SNe Ia in the solar neighborhood, though it should be said that these two assumed characteristics are highly speculative at present. These proposed SNe are denoted SNe Ia' in this report.

In accordance with increased occurrence of SNe Ia', if such a situation can be expected, a signature of SNe Ia' should be observable in the resultant elemental abundance patterns of the ISM. The most plausible signatures are (i) the enhancement of Ge and Ga, which are synthesized in SNe Ia' with highest production rates, and (ii) a significant enrichment in Fe during very early epochs, as described in the following.

\section{Chemical Evolution of a Galaxy at $z=2.626$}

\citet{Prochaska2_03} have opened a new window for the study of chemical evolution of distant galaxies through the use of a set of detailed elemental abundances. They have obtained abundances for about 25 elements in a DLA galaxy at $z=2.626$, although most of the elements including Fe suffer from a large uncertainty due to the estimate of dust corrections.  Since the [X/Fe] ratios for several elements act as good indicators of chemical evolution, it is useful to derive the Fe abundance in this galaxy. The derivation is performed in this study using the [Zn/Fe] ratio because zinc is an undepleted element and the nearly constant [Zn/Fe] ratio of [Fe/H] \gtsim--2.5 for solar neighborhood stars implies that the nucleosynthesis [Zn/Fe] ratio is always the same ($\sim$ 0.1). The [Zn/Fe] ratio of the ISM is therefore unrelated to the star formation histories of galaxies. Taking the possibility of a slight enhancement of Zn in this galaxy as described below, 
a [Zn/Fe] ratio of $0.15$ is assumed here, resulting in [Fe/H]~$=-0.86$ from the [Zn/H] ratio of $-0.71$ for this galaxy. The [X/Fe] ratios for the key elements are thus obtained as [O/Fe]=0.42, [Mg/Fe]=0.38, 
[Si/Fe]=0.25, [Mn/Fe]$<-0.39$, [Ga/Fe]$<0.11$, and [Ge/Fe]=0.24.

Two distinctive features can be seen in this result. First, the [$\alpha$/Fe] and [Mn/Fe] ratios clearly represent an SN II-like ratio. Second, [Ge/Fe] is significantly enhanced compared to the other ratios. The former feature implies that this galaxy is the progenitor of a massive elliptical galaxy, as already concluded by \citet{Prochaska2_03}. If so, high gas densities can be expected during the formation phase of this galaxy, which might result in high efficiency of production of close binary systems (see \S 1). The second feature can also be understood from this perspective. The frequency of SNe Ia' formation in this galaxy is estimated below, allowing the observed abundance pattern to be reproduced.

The relative frequencies of SNe Ia' and SNe II occurrence in this galaxy can be determined as follows. Let $N$ be the total number of SNe that have ever occurred in this galaxy until $z=2.626$, and let $w$ be the mass fraction of the synthesized elements that have been ejected from SNe and not locked up in low-mass stars. The total mass $M_X$ of element $X$ contained in the gas is then given by
$M_X = w_{\rm II}M_{X, \rm II} N_{\rm II}+w_{\rm Ia'}M_{X, \rm Ia'} N_{\rm Ia'} $, 
where $M_{X,\rm II}$ is the initial mass function (IMF) weighted average mass of element $X$ produced in SNe II, and $M_{X, \rm Ia'} $ is the mass from SNe Ia'.  For Fe, the mass $\sim 0.8$\ms ejected from SNe Ia' is roughly ten times larger than that from SNe II~ \citep[see][]{Tsujimoto_95}, and thus $M_{\rm Fe, Ia'}=10M_{\rm Fe, II}$ is assumed. In general, the values of $w_{\rm Ia}$ and $w_{\rm II}$ depend on the star formation history according to the production of heavy elements on different timescales by SNe Ia and SNe II~\citep{Tsujimoto_95}. In this case, a minor difference in timescales between SNe Ia' and SNe II would result in $w_{\rm Ia'}=w_{\rm II}$, leading to the equation, $M_{\rm Fe}=w_{\rm II} M_{\rm Fe, II}(N_{\rm II}+10N_{\rm Ia'})$.

For other elements, we utilize the fact that information on the nucleosynthesis [X/Fe]$_{\rm II}$ ratios for average SNe II yields can be derived from abundance data for metal-poor stars.  In other words, the observed [X/Fe] of metal-poor stars essentially corresponds to [X/Fe]$_{\rm II}$. Moreover, as already mentioned in \S 2, the [X/Fe]$_{\rm Ia'}$ ratios for SNe Ia' yields are given by two anomalous  stars. Using these results, 
we obtain $M_{X, \rm II}=(X/{\rm Fe})_\odot M_{\rm Fe, II} 10^{\rm [X/Fe]_{\rm II}}$ and\\ $M_{X, \rm Ia'}=10(X/{\rm Fe})_\odot M_{\rm Fe, II} 10^{\rm [X/Fe]_{\rm Ia'}}$, which gives the result, 
$M_X=w_{\rm II} M_{\rm Fe, II}(X/{\rm Fe})_\odot(10^{\rm [X/Fe]_{\rm II}}N_{\rm II}+10\times 10^{\rm [X/Fe]_{\rm Ia'}}N_{\rm Ia'})$.
Thus, the [X/Fe] ratio in the gas can be expressed in terms of $r=N_{\rm Ia'}/N_{\rm II}$ as follows,
\begin{equation}
{\rm [X/Fe]}=\log\frac{10^{\rm [X/Fe]_{II}}+10^{\rm [X/Fe]_{Ia'}+1}r}{1+10r} \ \ \ .
\end{equation}
This equation is identical to the formulations derived by \citet{Iwamoto_99} and \citet{Ivans_03}, and gives the frequency ratio $r$, which can then be used to reproduce the observed [Ge/Fe] ratio in this galaxy. The observed [Ge/Fe] ratios of metal-poor stars at [Fe/H] \ltsim $-2$ should represent the nucleosynthesis [Ge/Fe]$_{\rm II}$. A few metal-poor stars with observed [Ge/Fe] at [Fe/H]$<-2$ have been reported~ \citep{Sneden_98, Cowan_02}, including HD115444 with ([Ge/Fe],[Fe/H])~=~($-0.75$,$-2.77$), HD122563  with ($-0.80$,$-2.71$), and BD+17$^\circ$3248 with ($-0.72$,$-2.09$). Thus,  [Ge/Fe]$_{\rm II}=-0.75$ is adopted as an average value for metal-poor stars. For SNe Ia', [Ge/Fe]$_{\rm Ia'}$=$+1.75$ is used considering the [Ga/Fe] data for CS22966-043, assuming [Ga/Ge]~$=0$, which is expected  at least for SNe II nucleosynthesis from the [Ga/Fe] ratio of $-0.87$ for CS22941-012 \citep{Ivans_03} similar to [Ge/Fe] for above stars. These values in conjunction with [Ge/Fe]~$=0.24$ for this galaxy yield $r=0.003$.

It should be noted that the obtained frequency of SNe Ia' occurrence is very small, as is obvious in comparison with $N_{\rm Ia}/N_{\rm II} \sim 0.15-0.25$ for the solar neighborhood~\citep{Tsujimoto_95, Pagel_95, Ivans_03}. Such a small contribution from SNe Ia' will have little impact on abundance ratios other than those of Ga and Ge.  The yield ratios of [Mg/Fe]$_{\rm II}=0.4$ and [Mg/Fe]$_{\rm Ia'}=-0.65$ return a change of $\Delta$[Mg/Fe]$\sim -0.01$ from [Mg/Fe]$_{\rm II}$. A change of $\Delta$[Mn/Fe]$\sim+0.06$ is also obtained from [Mn/Fe]$_{\rm II}=-0.4$ and [Mn/Fe]$_{\rm Ia'}=0.4$. These  reproduce the observed SNe II-like abundance pattern. In the same way, the yield ratios of [Zn/Fe]$_{\rm II}=0.08$ and [Zn/Fe]$_{\rm Ia'}=1.0$ result in a change of $\Delta$[Zn/Fe]~$=+0.08$.

\section{Chemical Evolution of QSOs at $z \sim 6$}

\begin{figure}[ht]
\vspace{-2.5cm}
\begin{center}
\includegraphics[width=8cm,clip=true]{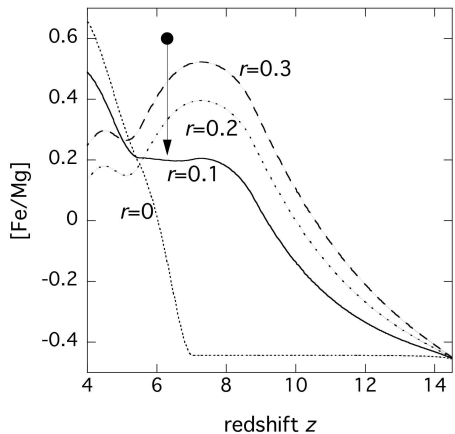}
\end{center}
\vspace{-2cm}
\caption{Evolution of [Fe/Mg] ratio in QSOs against redshift. Solid, dashed, long-dashed, and dotted curves denote models with $r=$~0.1, 0.2, 0.3, and 0, respectively. 
In the former three cases, $(N_{\rm Ia}+N_{\rm Ia'})/N_{\rm II}=0.2$ is fixed.  A cosmology of $H_0=71$~km~s$^{-1}$~Mpc$^{-1}$, $\Omega_\Lambda =0.73$, and $\Omega_m =0.27$ is adopted, and the formation epoch is set at $z=15$. The observational point is derived through conversion of the FeII/MgII line ratio~\citep{Maiolino_03} to the abundance [Fe/Mg] ratio (see text), and is indicated by an arrow taking into account uncertainty in microturbulence~\citep{Verner_03}.
 }
\end{figure}

The high metallicities (e.g., Hamann \& Ferland 1992, 1993) and high infrared luminosities \citep{Solomon_03} of high-$z$ QSOs are expected to be associated with a massive starburst. Therefore, an extremely high frequency of SNe Ia' occurrence would be expected in these QSOs because of the high gas density that triggered the starburst. The essential signature of enrichment by SNe Ia' is a high Fe abundance with a very short timescale, and in this case we have recognized twice the FeII/MgII line ratio of low-$z$ QSOs at $z \sim 6$. \citet{Verner_03} derived FeII/MgII ratios as functions of Fe abundance and microturbulence. According to their results, FeII/MgII~$= 8.65$ at $z \sim 6.3$~\citep{Maiolino_03} roughly corresponds to the Fe/Mg abundance ratio of $\sim 4 \times {\rm (Fe/Mg)}_\odot$ if microturbulence is similar to that of low-$z$ QSOs. Such a significant Fe enrichment is demanded to be achieved within less than 1~Gyr. 

Chemical evolution models incorporating this contribution from SNe Ia' can then be constructed in an attempt to reproduce the observations. The basic ingredients of the models are described in \citet{Yoshii_98}. The star formation rate (SFR) is assumed to be proportional to the gas density, with an SFR coefficient of 12.4~Gyr$^{-1}$~\citep{Arimoto_87}. This value yields a timescale for gas consumption of $\sim 10^8$~yr. Salpeter's IMF is adopted. The lifetime $t_{\rm Ia}$ of normal SNe Ia is set to be $t_{\rm Ia}=0.5-3$~Gyr, and is determined by the chemical evolution in the solar neighborhood~\citep{Yoshii_96}. However, there is a possibility that high-density environments will affect the timescale of these SNe Ia, in the sense that it would be made shorter \citep{Matteucci_01}. 
For SNe Ia',  a value of $t_{\rm Ia'}=10^7-3\times10^8$~yr is assumed. The upper bound for the lifetime of SNe Ia' is set so as to achieve a high Fe abundance, and is also required to be less than several  $10^8$~yrs as lifetimes that exceed the timescale for the formation of massive elliptical galaxies would change the abundance pattern of ISM in a galaxy as previously discussed, resulting in a pattern that is incompatible with the observations. 

The frequency of SNe Ia' occurrence is estimated based on the following considerations. The inferred SFR of a QSO (e.g., a Cloverleaf QSO at $z=2.56$) is about 30 times greater than that of starburst galaxies~\citep{Solomon_03}. If we regard starburst galaxies as the precursors of massive elliptical galaxies, the gas density $\rho$ during the QSO formation phase could be about 30 times larger than that of a galaxy at $z=2.626$, leading to $r \sim 0.1$, as deduced from the approximate relation that the number of close binaries $N \propto$ (encounter rate)$^{0.74} \propto (\rho^{1.5})^{0.74} \propto \rho$~\citep{Pooley_03}. The derived frequency corresponds to about half that of SNe Ia in the solar neighborhood. Figure 1 shows the evolution of [Fe/Mg] with redshift for four cases of different $r$ (0, 0.1, 0.2, and 0.3). The models with an occurrence frequency of SNe Ia' comparable to that of SNe Ia in the solar neighborhood can satisfactorily reproduce the observations. Without a contribution from SNe Ia' (the case with $r$=0), more than 1~Gyr is required to achieve the observed Fe/Mg ratio. 

\section{Conclusions}

A new class of supernovae, most likely SNe Ia, is proposed as an end result of close binary evolution. Although these SNe are rare in the solar neighborhood and nearby galaxies, fossil imprints of these SNe have been recently detected in elemental abundances on the surface of two stars in the solar neighborhood, without sign of these SNe in the present ISM. In high-$z$ galaxies that undergo massive star formation, these SNe may represent a significant population, being dependent on high prevailing gas densities and the resultant increased formation of close binary systems. As a result, high-$z$ galaxies exhibit unique elemental abundance patterns in the gas as observed, that is, enhanced [Ge (Ga)/Fe] ratios and high Fe with a very short timescale. These features cannot be explained by any combination of SNe known at present. 

The additional abundance studies for the DLA galaxies as performed by Prochaska  et al.  will lead to a more complete understanding of the nature of this new class of SNe. Further, the expected high occurrence frequency of these SNe in high-$z$ galaxies predicts that the scatter in the luminosity of  SNe Ia which act as standard candles becomes large in a high-$z$ universe owing to an additional contribution of these SNe which have somewhat brighter luminosity. It should be, however, noted that 
the scenario described in this {\it letter} is based on a number of highly speculative hypotheses. Any decision on the validity of the existence of these new SNe should await detailed calculations of nucleosynthesis in SNe, including the analysis of super-Chandraseker mass models for SNe Ia, as well as of the coalescence process of double WDs.

\acknowledgements

The author is particularly appreciative of the referee R. Gallino, and also the anonymous referees for useful  comments that helped improve this paper.

\end{document}